\documentclass[dvips]{arxstspdf}
\usepackage{dcolumn}
\usepackage{graphicx,flushend,stfloats}

\volume{25}
\issue{2}
\pubyear{2010}
\firstpage{145}
\lastpage{157}
\doi{10.1214/09-STS308}

\makeatletter
\newcommand{\Fdr}{\operatorname{Fdr}}

\newcolumntype{d}[1]{D{.}{.}{#1}}

\newcommand{\hFdr}{\widehat\Fdr}
\newcommand{\n}{\mathcal{N}}
\newcommand{\ind}{\stackrel{\mathrm{ind}}{\sim}}

\newcommand{\eqref}[1]{(\ref{#1})}

\newtheorem{lemma}{Lemma}

\makeatother

\begin{document}
\begin{frontmatter}

\title{The Future of Indirect Evidence{\protect\thanksref{T2}}}
\runtitle{The Future of Indirect Evidence}
\relateddoi{T2}{Discussed in \doi{10.1214/10-STS308A},
\doi{10.1214/10-STS308B} and
\doi{10.1214/10-STS308C};
rejoinder at
\doi{10.1214/10-STS308REJ}.}

\begin{aug}
\author{\fnms{Bradley} \snm{Efron}\ead[label=e1]{brad@stat.stanford.edu}\corref{}}
\runauthor{B. Efron}

\affiliation{Stanford University}

\address{Bradley Efron is Professor, Department of Statistics, Stanford
    University, Stanford, California 94305, USA.}

\end{aug}

\begin{abstract}
Familiar statistical tests and estimates are obtained by the direct
  observation of cases of interest: a clinical trial of a new drug,
  for instance, will compare the drug's effects on a relevant set of
  patients and controls. Sometimes, though, \textit{indirect evidence}
  may be temptingly available, perhaps the results of previous trials
  on closely related drugs. Very roughly speaking, the difference
  between direct and indirect statistical evidence marks the boundary
  between frequentist and Bayesian thinking. Twentieth-century
  statistical practice focused heavily on direct evidence, on the
  grounds of superior objectivity. Now, however, new scientific
  devices such as microarrays routinely produce enormous data sets
  involving thousands of related situations, where indirect evidence
  seems too important to ignore. Empirical Bayes methodology offers an
  attractive direct/indirect compromise. There is already some
  evidence of a shift toward a less rigid standard of statistical
  objectivity that allows better use of indirect evidence. This
  article is basically the text of a recent talk featuring some
  examples from current practice, with a little bit of futuristic
  speculation.
\end{abstract}


\end{frontmatter}

\section{Introduction}\label{sec1}

This article is the text of a talk I gave twice in 2009, at the
Objective Bayes Conference at Wharton, and at the  Joint
Statistical Meetings in Washington, DC. Well, not quite the text. The printed page gives
me a chance to repair a couple of the more gaping omissions in the
verbal presentation, without violating its rule of avoiding almost all
mathematical technicalities.

Basically, however, I'll stick to the text, which was a broad-brush
view of some recent trends in statistical applications---their
rapidly increasing size and complexity---that are impinging on
statistical theory, both frequentist and Bayesian. An OpEd piece on
``practical philosophy'' might be a good description of what I was
aiming for. Most of the talk (as I'll refer to this from now on) uses
simple examples, including some of my old favorites, to get at the
main ideas.  There is no attempt at careful referencing, just a short
list of directly relevant sources mentioned at the end.

\begin{figure*}

\includegraphics{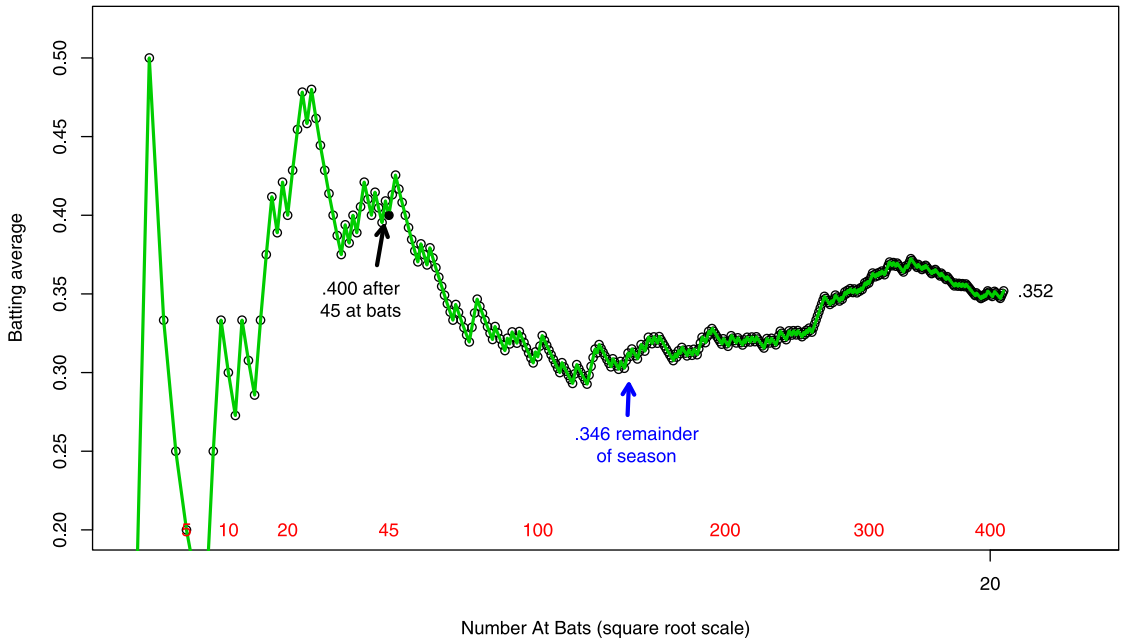}

\caption{Roberto Clemente's batting averages over the 1970 baseball
  season (\hspace*{0.6pt}partially simulated\hspace*{0.3pt}). After 45 tries he had 18 hits for a
  batting average of $18/45=0.400$; his average in the remainder of the
  season was $127/367=0.346$.}\label{fig1}
\end{figure*}

I should warn you that the talk is organized more historically than
logically. It starts with a few examples of frequentist, Bayesian and
empirical Bayesian analysis, all bearing on ``indirect evidence,'' my
catch-all term for useful information that isn't of obvious direct
application to a question of interest. This is by way of a long
build-up to my main point concerning the torrent of indirect evidence
uncorked by modern scientific technologies such as the microarray. It
is fair to say that we are living in a new era of statistical
applications, one that is putting pressure on traditional Bayesian and
frequentist methodologies. Toward the end of the talk I'll try to
demonstrate some of the pitfalls and opportunities of the new era,
finishing, as the title promises, with a few words about the future.

\section{Direct Statistical Evidence}\label{sec2}

A statistical argument, at least in popular parlance, is one in which
many small pieces of evidence, often contradictory, are amassed to
produce an overall conclusion. A familiar and important example is the
clinical trial of a promising new drug. We don't expect the drug to
work on every patient, or for every placebo-receiving patient to fail,
but perhaps, overall, the new drug will perform ``significantly''
better.

The clinical trial is collecting \textit{direct statistical evidence},
where each bit of data, a patient's success or failure, directly bears
on the question of interest. Direct evidence, interpreted by
frequentist methods, has been the prevalent mode of statistical
application during the past century. It is strongly connected with the
idea of scientific objectivity, which accounts, I believe, for the
dominance of frequentism in scientific reporting.

Figure \ref{fig1} concerns an example of direct statistical evidence, taken
from the sports pages of 1970. We are following the star baseball
player Roberto Clemente through his 1970 season. His batting average,
number of successes (``hits'') over number of tries (``at bats'')
fluctuates wildly at first but settles down as the season progresses.
After 45 tries he has 18 hits, for a batting average of $18/45=0.400$
or ``four hundred'' in baseball terminology. The remainder of the
season is slightly less successful, with 127 hits out of 367 at bats
for a batting average of $0.346=127/367$, giving Clemente a full season
average of $0.352$.\footnote{These numbers are accurate, but I have to
  admit to simulating the rest of the figure by randomly dispersing
  his 18 hits over the first 45 tries, and similarly for the last 127
  hits.} This is a classic frequentist estimate: direct statistical
evidence for Clemente's 1970 batting ability.

In contentious areas such as drug efficacy, the desire for direct
evidence can be overpowering. A clinical trial often has three
arms: placebo, single dose of new drug, and double dose. Even if the
double dose/placebo comparison yields strongly significant results in
favor of the new drug, a not-quite significant result for the
single dose/placebo comparison, say $p$-value $0.07$, will not be enough
to earn FDA approval. The single dose \textit{by itself} must prove
its worth.

My own feeling at this point would be that the single dose is very
likely to be vindicated in any subsequent testing. The strong result
for the double dose adds \textit{indirect evidence} to the direct,
nearly significant, single dose outcome. As the talk's title suggests,
indirect statistical evidence is the focus of interest here. My main
point, which will take a while to unfold, is that current scientific
trends are producing larger and more complex data sets in which
indirect evidence has to be accounted for: and these trends will force
some re-thinking of both frequentist and Bayesian practices.

\section{Bayesian Inference}\label{sec3}

I was having coffee with a physicist friend and her husband who,
thanks to the miracle of sonograms, knew they were due to have twin
boys. Without warning, the mother-to-be asked me what was the
probability her twins would be identical rather than
fraternal. Stalling for time, I asked if the doctor had given her any
further information. ``Yes, he said the proportion of identical twins
is one-third.'' (I checked later with an epidemiology colleague who
confirmed this estimate.)

Thomas Bayes, 18th-century non-conformist English minister, would have
died in vain if I didn't use his rule to answer the physicist mom. In
this case the prior odds
\[
\frac{\Pr\{\mathrm{Identical}\}}{\Pr\{\mathrm{Fraternal}\}}=\frac{1/3}{2/3} = \dfrac12
\]
favor fraternal. However the likelihood ratio, the current evidence
from the sonogram, favors identical,
\[
\frac{\Pr\{\mathrm{Twin\ boys|Identical}\}}{\Pr\{\mathrm{Twin\ boys|Fraternal}\}}=2,
\]
since identical twins are always the same sex while fraternal twins
are of differing sexes half the time.

Bayes rule, published posthumously in 1763, is a rule for combining
evidence from different sources. In this case it says that the
posterior odds of identical to fraternal is obtained by simple
multiplication.
\begin{eqnarray*}
\mbox{Posterior odds }&=&\mbox{(Prior odds)} \cdot \mbox{(Likelihood ratio)}\\
&=&\tfrac12\cdot2=1.
\end{eqnarray*}
So my answer to the physicists was ``$50/50$,'' equal chances of
identical or fraternal. (This sounded like pure guessing to them; I
would have gotten a lot more respect with ``$60/40$.'')

Bayes rule is a landmark achievement. It was the first breakthrough in
scientific logic since the Greeks and the beginning of statistical
inference as a serious mathematical subject. From the point of view of
this talk, it also marked the formal introduction of indirect evidence
into statistical learning.

Both Clemente and the physicists are learning from
experience. Clemente is learning directly from his own experience, in
a strict frequentist manner. The physicists are learning from their
own experience (the sonogram), but also indirectly from the
experience of others: that one-third/two-thirds prior odds is based on
perhaps millions of previous twin births, mostly not of the
physicists' ``twin boys'' situation. Another way to state Bayes rule
is as a device for filtering out and using the relevant portions of
past experiences.

All statisticians, or almost all of them, enjoy Bayes rule, but only a
minority make much use of it. Learning \textit{only} from direct
experience is a dominant feature of contemporary applied statistics,
connected, as I said, with notions of scientific objectivity. A
fundamental Bayesian difficulty is that well-founded prior
distributions, like the twins one-thirds/two-thirds, are rare in
scientific practice. Much of 20th-century Bayesian theory concerned
subjective prior distributions, which are not very convincing in
contentious areas such as drug trials.

The holy grail of statistical theory is to use the experience of
others without the need for subjective prior distributions: in
L. J.~Savage's words, to enjoy the Bayesian omelette without breaking
the Bayesian eggs. I am going to argue that this grail has grown
holier, and more pressing, in the 21st century. First though I wanted
to say something about frequentist use of indirect information.

\section{Regression Models}\label{sec4}

Bayesians have an advantage but not a monopoly on the use of indirect
evidence. Regression models provide an officially
sanctioned\footnote{Sanctioned, though not universally accepted as
  fully relevant, as the three-arm drug example showed.} frequentist
mechanism for incorporating the experience of others.
%
\begin{figure*}

\includegraphics{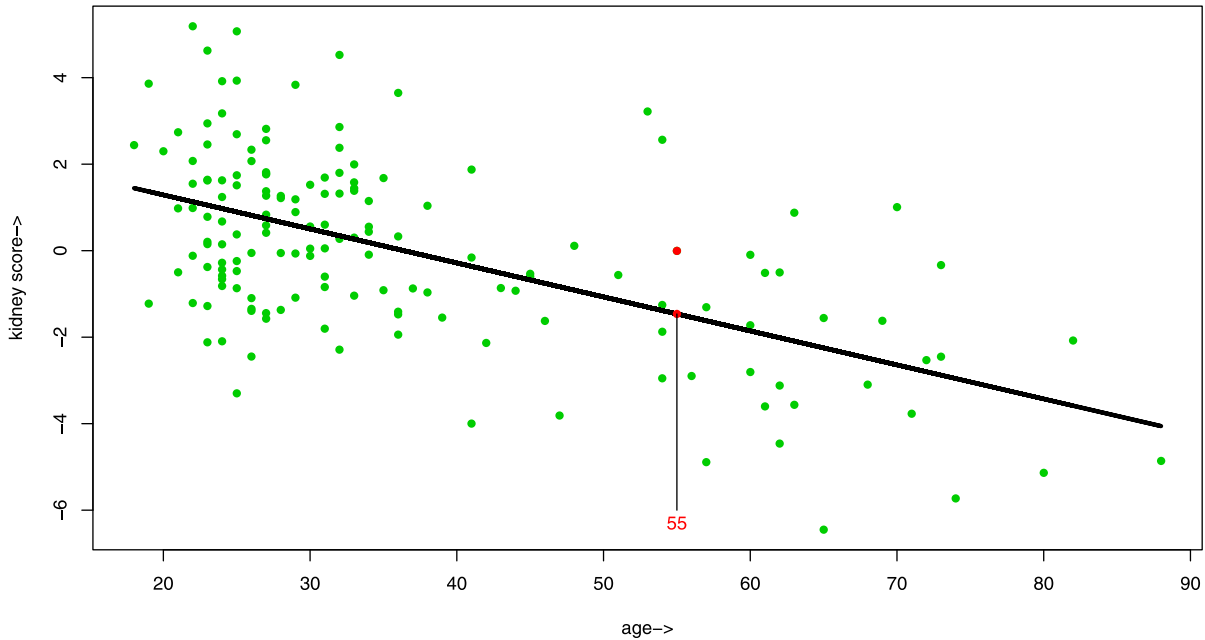}

\caption{Kidney function plotted versus age for 157 healthy volunteers
  from the nephrology laboratory of Dr.~Brian Myers. The least squares
  regression line has a strong downward slope. A new donor age 55 has
  appeared, and we need to predict his kidney score.}\label{fig2}
\end{figure*}

Figure \ref{fig2} concerns an example from Dr.~Brian Myers' Stanford
nephrology laboratory: 157 healthy volunteers have had their kidney
function evaluated by a somewhat arduous series of tests. An overall
kidney score, higher numbers better, is plotted versus the volunteer's
age, illustrating a decline in function among the older
subjects. (Kidney donation was once limited to volunteers less than 60
years old.) The decline is emphasized by the downward slope of the
least squares regression line.

A potential new donor, age 55, has appeared but it is not practical to
evaluate his kidney function by the arduous testing procedure. How
good are his kidneys? As far as direct evidence is concerned, only one
of the 157 volunteers was 55, and he had score $-0.01$. Most
statisticians would prefer the estimate obtained from the height at
age 55 of the least square line, $-1.46$. In Tukey's evocative
language, we are ``borrowing strength'' from the 156 volunteers who
are not age~55.

Borrowing strength is a clear use of indirect evidence, but invoked
differently than through Bayes theorem. Now every individual is
adjusted to fit the case of interest; in effect the regression model
allows us to adjust each volunteer to age 55. Linear model theory
permits a direct frequentist analysis of the entire least squares
fitting process, but that shouldn't conceal the indirect nature of its
application to individual cases.

One response of the statistical community to the onslaught of
increasingly large and complex data sets has been to extend the reach
of regression models: LARS, lasso, boosting, bagging, CART and
projection pursuit being a few of the ambitious new data-mining
algorithms. Every self-respecting sports program now has its own
simplified data-mining program, producing statements like ``Jones has
only 3 hits in 16 tries versus Pettitte.'' This is direct evidence run
amok. Regression models seem to be considered beyond the sporting
public's sophistication, but indirect evidence is everywhere in the
sports world, as I want to discuss next.

\section{James--Stein Estimation}\label{sec5}

Early in the 1970 baseball season, Carl Morris collected the batting
average data shown in the second column of Table \ref{table1}. Each of the
18 players had batted 45 times (they were all of those who had done
so) with varying degrees of success. Clemente, as shown in Figure \ref{fig1},
had hit successfully 18 of the 45 times, for an observed average of
$0.400=18/45$. Near the bottom of the table, Thurman Munson, another
star player, had only 8 hits; observed average $8/45=0.178$. The grand
average of the 18 players at that point was $0.265$.

\begin{table}[b]
\tabcolsep=0pt
\caption{Batting averages for 18 major league players early in the
  1970 season (``Observed'') and their averages for the
  remainder of the season (``Truth''). Also the James--Stein predictions}\label{table1}
\begin{tabular*}{\columnwidth}{@{\extracolsep{\fill}}lcccc@{}}
\hline
\textbf{Name}&\textbf{Hits$\bolds{/}$AB}&\textbf{Observed}&\textbf{``Truth''}& \textbf{James--Stein}\\
\hline
\hphantom{0}1. Clemente&$18/45$&0.400&0.346&0.294\\
\hphantom{0}2. F.~Robinson&$17/45$&0.378&0.298&0.289\\
\hphantom{0}3. F.~Howard  &$16/45$&0.356&0.276&0.285\\
\hphantom{0}4. Johnstone  &$15/45$&0.333&0.222&0.280\\
\quad\qquad$\vdots$&$\vdots$&$\vdots$&$\vdots$&$\vdots$\\
14. Petrocelli   &$10/45$&0.222&0.264&0.256\\
15. E.~Rodriguez &$10/45$&0.222&0.226&0.256\\
16. Campaneris  &$\hphantom{0}9/45$&0.200&0.286&0.252\\
17. Munson&$\hphantom{0}8/45$&0.178&0.316&0.247\\
18. Alvis&$\hphantom{0}7/45$&0.156&0.200&0.242\\[5pt]
Grand average&&0.265&0.265&0.265\\
\hline
\end{tabular*}
\end{table}

Only about one-tenth of the season had elapsed, and Morris considered
predicting each player's subsequent batting average during the
remainder of 1970. Since the players bat independently of each other---Clemente's successes don't help Munson, nor vice versa---it
seems there is no alternative to using the observed averages, at least
not without employing more baseball background knowledge.

However, that is not true. The \textit{James--Stein estimates} in the
last column of the table are functions of the observed averages,
obtained by shrinking them a certain amount of the way toward the
grand average $0.265$, as described next. By the end of the 1970
season, Morris could see the ``truth,'' the players' averages over the
remainder of the season. If prediction error is measured by total
squared discrepancy from the truth, then James--Stein wins handsomely:
its total squared prediction error was less than one-third of that for
the observed averages. This wasn't a matter of luck, as we will see.

Suppose each player has a true expectation $\mu_i$ and an observed
average $x_i$, following the model
\begin{equation}\label{1}
\mu_i\sim\n(M,A)\quad\mbox{and}\quad x_i|\mu_i\sim\n(\mu_i,\sigma_0^2)
\end{equation}
for $i=1,2,\dots,N=18$. Here $M$ and $A$ are mean and variance
hyper-parameters that determine the Bayesian prior distribution;
$\mu_i$ can be thought of as the ``truth'' in Table \ref{table1}, $x_i$ as
the observed average, and $\sigma_0^2$ as its approximate binomial
variance $0.265\cdot(1-0.265)/45$. (I won't worry about the fact that
$x_i$ is binomial rather than perfectly normal.)

The posterior expectation of $\mu_i$ given $x_i$, which is the Bayes
estimator under squared error loss, is
\begin{eqnarray}\label{2}
\hat\mu_i^{(\mathrm{Bayes})}=M+B(x_i-M)\nonumber\\[-8pt]\\[-8pt]
\eqntext{\mbox{where }B=\dfrac{A}{A+\sigma_0^2}.}
\end{eqnarray}
If $A=\sigma_0^2$, for example, Bayes rule shrinks each observed
average $x_i$ half way toward the prior mean $M$. Using Bayes rule
reduces the total squared error of prediction, compared to using the
obvious estimates $x_i$, by a factor of $1-B$. This is a 50\% savings
if $A=\sigma_0^2$, and more if the prior variance $A$ is less than
$\sigma_0^2$.

Baseball experts might know accurate values for $M$ and $A$, or $M$
and $B$, but we are not assuming expert prior knowledge here. The
James--Stein estimator can be motivated quite simply: unbiased
estimates $\hat{M}$ and $\hat{B}$ are obtained from the vector of
observations $\mathbf{x}=(x_1,x_2,\dots,x_N)$ (e.g., $\hat{M}=\bar{x}$ the
grand average) and substituted into formula \eqref{2}. In Herbert
Robbins' apt terminology, James--Stein is an \textit{empirical Bayes}
estimator. It doesn't perform as well as the actual Bayes estimate
\eqref{2}, but under model \eqref{1} the penalty is surprisingly
small.

All of this seems interesting enough, but a skeptic might ask where
the normal prior distributions $\mu_i\ind\n(M,A)$ in \eqref{1} are
coming from. In fact, James and Stein didn't use normal priors, or any
priors at all, in their derivation. Instead they proved the following
frequentist theorem.
\begin{thm}[(1956)]\label{thm1}
  If $x_i\sim\n(\mu_i,\sigma_0^2)$ independently for $i=1,2,\dots,N,\
  N\geq4$, then the James--Stein estimator \emph{always} beats the
  obvious estimator $x_i$ in terms of expected total squared
  estimation error.
\end{thm}

This is the single most striking result of post-World War II
statistical theory. It is sometimes called\footnote{Willard James was
  Charles Stein's graduate student. Stein had shown earlier that
  another, less well-motivated, estimator dominated the obvious rule.}
\textit{Stein's paradox} for it says that Clemente's good performance
\textit{does} increase our estimate for Munson (e.g., by increasing
$\hat{M}=\bar{x}$) and vice versa, even though they succeed or fail
independently. In addition to the direct evidence of each player's
batting average, we gain indirect evidence from the other 17 averages.

James--Stein estimation is not an unmitigated blessing. Low total
squared error can conceal poor performance on genuinely unusual
cases. Baseball fans know from past experience that Clemente was an
unusually good hitter, who is learning too much from the experience of
others by being included in a cohort of less-talented players. I'll
call this the \textit{Clemente problem} in what follows.

\section{Large-Scale Multiple Inference}\label{sec6}

All of this is a preface, and one that could have been written 50
years ago, to what I am really interested in talking about
here. Large-scale multiple inference, in which thousands of
statistical problems are considered at once, has become a fact of life
for 21st-century statisticians. There is just too much indirect
evidence to ignore in such situations. Coming to grips with our new,
more intense, scientific environment is a major enterprise for the
statistical community, and one that is already affecting both theory
and practice.

Rupert Miller's book \textit{Simultaneous Statistical Inference}
appeared in 1966, lucidly summarizing the post-war boom in
multiple-testing theory. The book is overwhelmingly frequentist, aimed
mainly at the control of type I error, and concerned with the
simultaneous analysis of between 2 and perhaps 10 testing problems.
Microarray technology introduced in the 1990s dramatically raised
the ante: number of problems $N$ now easily exceeds 10,000; ``SNP
chips'' have $N=500\mbox{,}000+$, and imaging devices reach higher still.

\begin{figure*}

\includegraphics{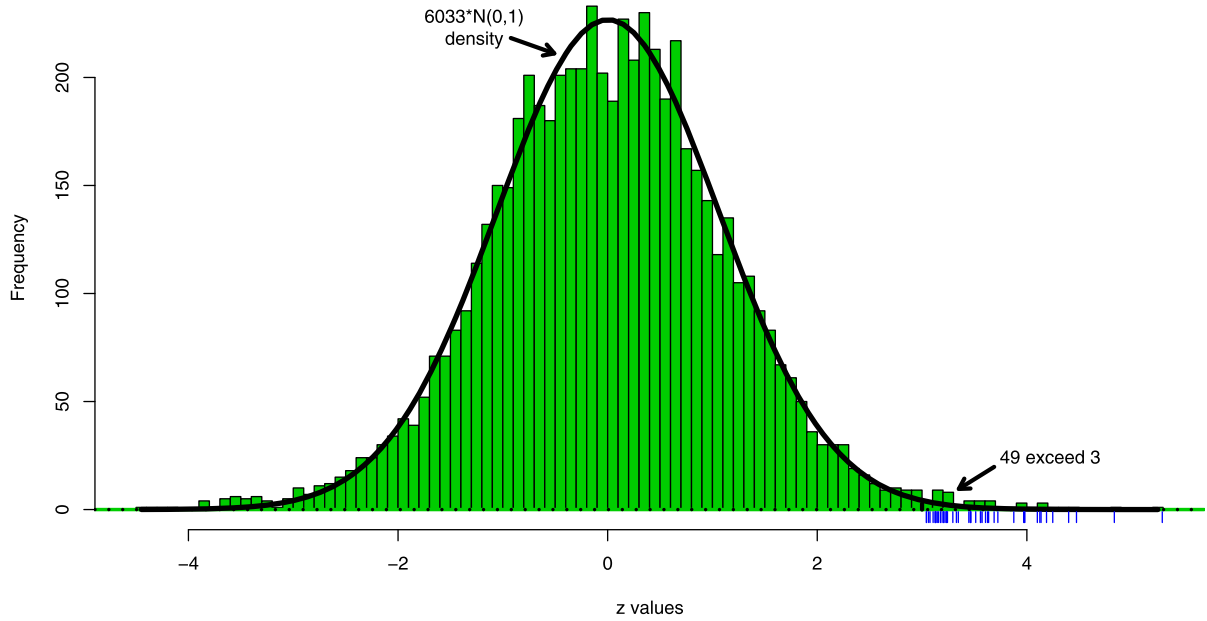}

\caption{Histogram of $N=6033 z$-values from the prostate cancer
  study compared with the theoretical null density that would apply
  if all the genes were uninteresting. Hash marks indicate the 49
  $z$-values exceeding $3.0$.}\label{fig3}
\end{figure*}

Figure \ref{fig3} concerns a microarray study in which the researchers were
on a fishing expedition to find genes involved in the development of
prostate cancer: 102 men, 50 healthy controls and 52 prostate cancer
patients, each had expression levels for $N=6033$ genes measured on
microarrays. The resulting data matrix had $N=6033$ rows, one for each
gene, and 102 columns, one for each man.

As a first step in looking for ``interesting'' genes, a two-sample
$t$-statistic $t_i$ comparing cancer patients with controls was
computed for each gene $i$, $i=1,2,\ldots,N$, and then converted to a
$z$-value
\begin{equation}\label{3}
z_i=\Phi^{-1}(F_{100}(t_i))
\end{equation}
with $\Phi$ and $F_{100}$ the c.d.f.'s of a standard normal and $t_{100}$
variate. Under the usual textbook conditions, $z_i$ will have a
standard normal distribution in the null (uninteresting) situation
where genetic expression levels are identically distributed for
controls and patients,
\begin{equation}\label{4}
H_0\dvtx z_i\sim\n(0,1).
\end{equation}

A histogram of the $N=6033\ z$-values appears in Figure \ref{fig3}. It is fit
reasonably well by the ``theoretical null'' curve that would apply if
all the genes followed \eqref{4}, except that there is an excess of
tail values, which might indicate some interesting ``non-null'' genes
responding differently in cancer and control subjects.

\begin{figure*}

\includegraphics{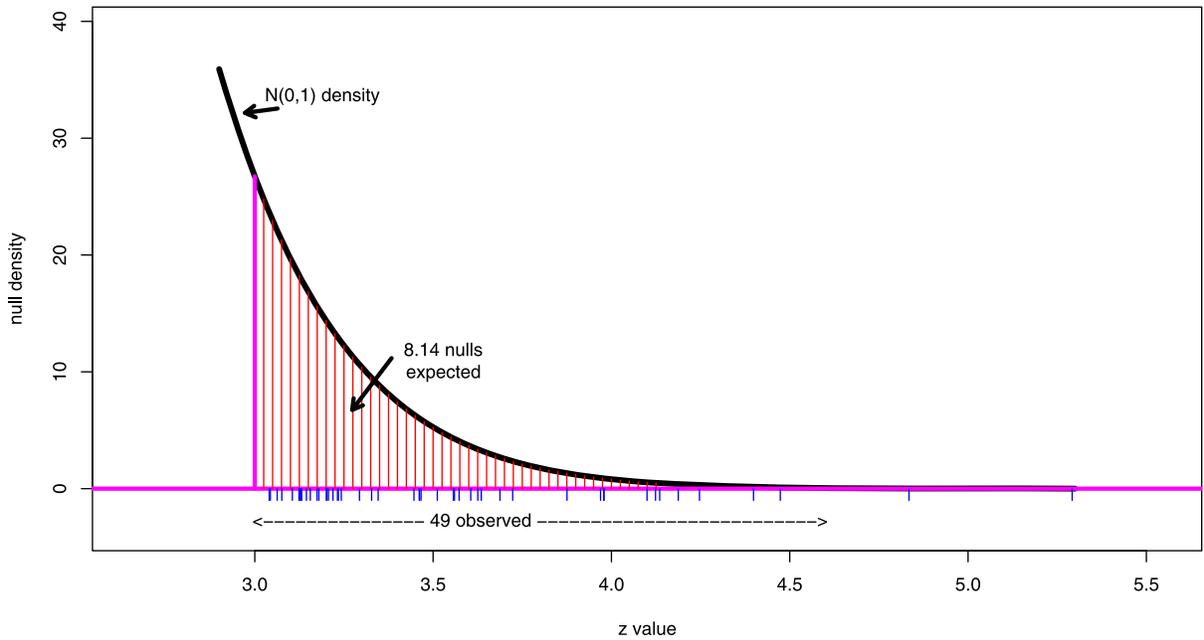}

\caption{Close-up of right tail of the prostate data $z$-value
  histogram; 49 $z_i$'s exceed $3.0$, compared to an expected number
  $8.14$ if all genes were null (\protect\ref{4}).}\label{fig4}
\end{figure*}

Here I will concentrate on the 49 genes having $z_i$ exceeding $3.0$,
as indicated by the hash marks. Figure~\ref{fig4} shows a close-up of the
right tail, where we notice that 49 is much greater than $8.14$, the
expected number of $z_i$'s exceeding $3.0$ under full null conditions.
The ratio is
\begin{equation}\label{5}
\hFdr(3.0)=\tfrac{8.14}{49}=\tfrac16.
\end{equation}
where Fdr stands for \textit{false discovery rate}, in Benjamini and
Hochberg's evocative terminology. Reporting the list of 49 back to the
investigators seems like a good bet if it only contains $1/6$ duds,
but can we believe that value?

Benjamini and Hochberg's (\citeyear{ben95}) paper answered the question with what I
consider the second most striking theorem of post-war statistics. For
any given cutoff point $c$ let $N(c)$ be the number of $z_i$'s
observed to exceed $c$, $E_0(c)$ the expected number exceeding $c$ if
all genes are null \eqref{4}, and
\begin{equation}
\hFdr(c)=E_0(c)/N(c).
\label{6}
\end{equation}
[In \eqref{5}, $c=3.0$, $N(c)=49$, and $E_0(c)=8.14$.] Choose an Fdr
control value $q$ between 0 and 1 and let $c_q$ be the smallest value
of $c$ such that $\hFdr(c)\leq q$.
\begin{thm}
  If the $N$ $z$-values are independent of each other, then the rule
  that rejects the null hypothesis \eqref{4} for all cases having
  $z_i\geq c_q$ will make the expected proportion of false discoveries
  no greater than $q$.
\label{thm2}
\end{thm}

In the prostate data example, choosing $q=1/6$ gives $c_q=3.0$ and
yields a list of 49 presumably interesting genes. Assuming
independence\footnote{This isn't a bad assumption for the prostate data, but
  a dangerous one in general for microarray experiments. However,
  dependence usually has little effect on the theorem's conclusion. A
  more common choice of $q$ is $0.10$.}, the theorem says that the
expected proportion of actual null cases on the list is no greater
than $1/6$. That is a frequentist expectation, \citeauthor{ben95}
like James and Stein having worked frequentistically, but once again
there is an instructive Bayesian interpretation.

A very simple Bayes  model for simultaneous hypothesis testing, the
\textit{two-groups model}, assumes that each gene has prior
probability $p_0$ or $p_1=1-p_0$ of being null or non-null, with
corresponding $z$-value density $f_0(z)$ or $f_1(z)$:
\begin{equation}\label{7}
\mbox{Prior probability}\cases{p_0,\cr p_1,}\quad z_i\sim\cases{f_0(z),\cr f_1(z).}
\end{equation}
Let $F_0(z)$ and $F_1(z)$ be the right-sided c.d.f.'s (\textit{survival
  functions}) corresponding to $f_0$ and $f_1$, and $F(z)$ their
mixture,
\begin{equation}
F(z)=p_0F_0(z)+p_1F_1(z).
\label{8}
\end{equation}
Applying Bayes theorem shows that the true false discovery rate is
\begin{eqnarray}\label{9}
\Fdr(c)&\equiv&\Pr\{\mbox{gene $i$ null}|z_i\geq
c\}\nonumber\\[-8pt]\\[-8pt]
&=&p_0F_0(c)/F(c).\nonumber
\end{eqnarray}
(Left-sided c.d.f.'s perform just as well, but it is convenient to work on
the right here.)

\begin{figure*}

\includegraphics{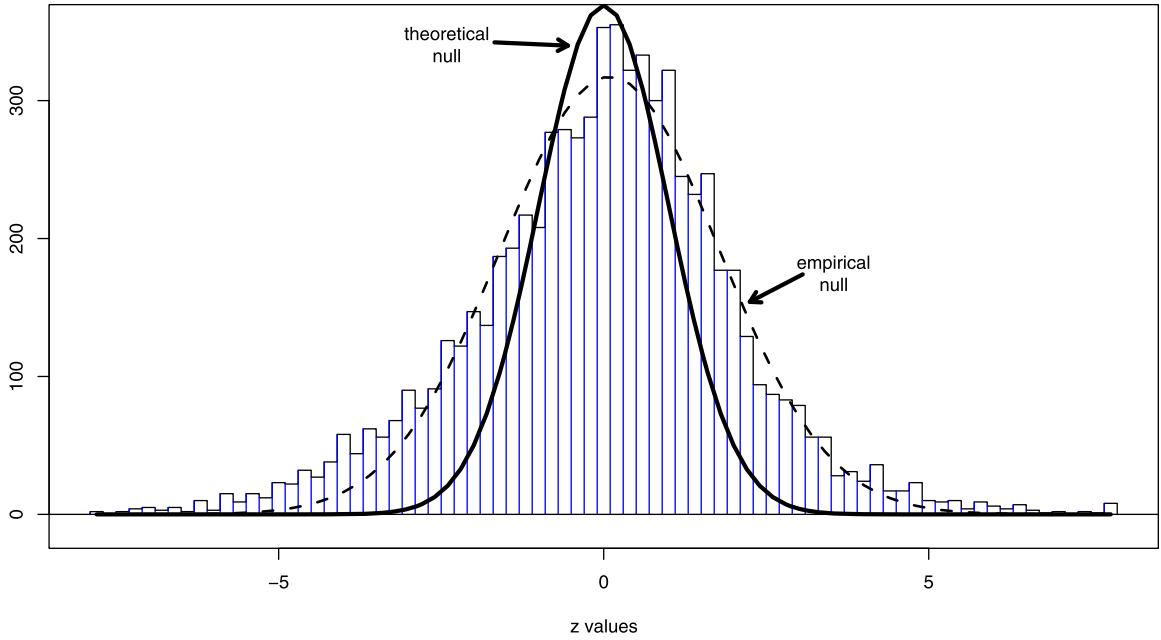}

\caption{Histogram of $z$-values for $N=7128$ genes in a microarray
  study comparing two types of leukemia. The $\n(0,1)$ theoretical
  null is much narrower than the histogram center; a normal fit to the
  central histogram height gives empirical null $\n(0.09,1.68^2)$. Both
  curves have been scaled by their respective estimates of $p_0$ in
  (\protect\ref{7}).}\label{fig5}
\end{figure*}

Of course we can't apply the Bayesian result \eqref{9} unless we know
$p_0$, $f_0$, and $f_1$ in \eqref{7}. Once again though, a simple
empirical Bayes estimate is available. Under the theoretical null
\eqref{4}, $F_0(z)=1-\Phi(z)$, the standard normal right-sided c.d.f.;
$p_0$ will usually be close to 1 in fishing expedition situations and
has little effect on $\Fdr(c)$. (Benjamini and Hochberg set $p_0=1$. It
can be estimated from the data, and I will take it as known here.)
That leaves the mixture c.d.f. $F(z)$ as the only unknown. But by
definition, all $N\ z_i$ values follow $F(z)$, so we can estimate it
by the empirical c.d.f. $\hat{F}(z)=\#\{z_i\geq z\}/N$, leading to the
empirical Bayes estimate of \eqref{9},
\begin{equation}
\hFdr(c)=p_0F_0(c)/\hat{F}(c).
\label{10}
\end{equation}

The two definitions of $\hFdr(c)$, \eqref{6} and \eqref{10}, are the
same since $E_0(c)=Np_0F_0(c)$ and $\hat{F}(c)=N(c)/N$. This means we
can restate \citeauthor{ben95}'s theorem in empirical Bayes terms: the
list of cases reported by BH($q$), the Benjamini--Hochberg-level $q$
rule, is essentially those cases having estimated posterior
probability of being null no greater than $q$.

The Benjamini--Hochberg algorithm clearly\break involves indirect
evidence. In this case, each $z$-value is learning from the other
$N-1$ values: if, say, only 10 instead of 49 $z$-values had exceeded
$3.0$, then $\hFdr(c)$ would equal $0.81$ (i.e., ``very likely null'')
so a gene with $z_i\geq3.0$ would now \textit{not} be reported as
non-null.

I have been pleasantly surprised at how quickly false discovery rate
control was accepted by statisticians and our clients. It is
fundamentally different from type I error control, the standard for
nearly a century, in its Bayesian aspect, its use of indirect
evidence, and in the fact that it provides an explicit
\textit{estimate} of nullness $\hFdr(z)$ rather than just a yes/no
decision.\footnote{Although one might consider $p$-values to provide such
  estimates in classical testing.}

\section{The Proper Use of Indirect Evidence}\label{sec7}

The false discovery rate story is a promising sign of our
profession's ability to embrace new methods for new problems. However,
in moving beyond the confines of classical statistics we are also
moving outside the wall of protection that a century of theory and
experience has erected against inferential error.

Within its proper venue, it is hard to go very wrong with a
frequentist analysis of direct evidence. I find it quite easy to go
wrong in large-scale data analyses. This section and the next offer a
couple of examples of the pitfalls yawning in the use of indirect
evidence. None of this is meant to be discouraging: difficulties are
what researchers thrive on, and I fully expect statisticians to
successfully navigate these new waters.

The results of another microarry experiment, this time concerning
leukemia, are summarized in Figure~\ref{fig5}. High-density oligonucleotide
microarrays provided expression levels on $N=7128$ genes for 72
patients, 45 with ALL (acute lymphoblastic leukemia) and 27 with AML
(acute myeloid leukemia), the latter having worse prognosis. Two-sample
$t$-statistics provided $z$-values $z_i$ for each gene, as with the
prostate study.

Figure \ref{fig5} shows that this time the center of the $z$-value histogram
does \textit{not} approximate a $\n(0,1)$ density. Instead, it is much
too wide: a maximum likelihood fit to central histogram heights gave
estimated proportion $p_0=0.93$ of null genes in the two-groups model
\eqref{7}, and an empirical null density estimate
\begin{equation}\label{11}
f_0(z)\sim\n(0.09,1.68^2),
\end{equation}
more than half again as wide as the $\n(0,1)$ theoretical null
\eqref{4}. The dashed curve shows \eqref{11} nicely following the
histogram height near the center while the estimated proportion of
non-null genes $p_1=1-p_0=0.07$ appear as heavy tails, noticeably on
the left.

\begin{figure*}[b]

\includegraphics{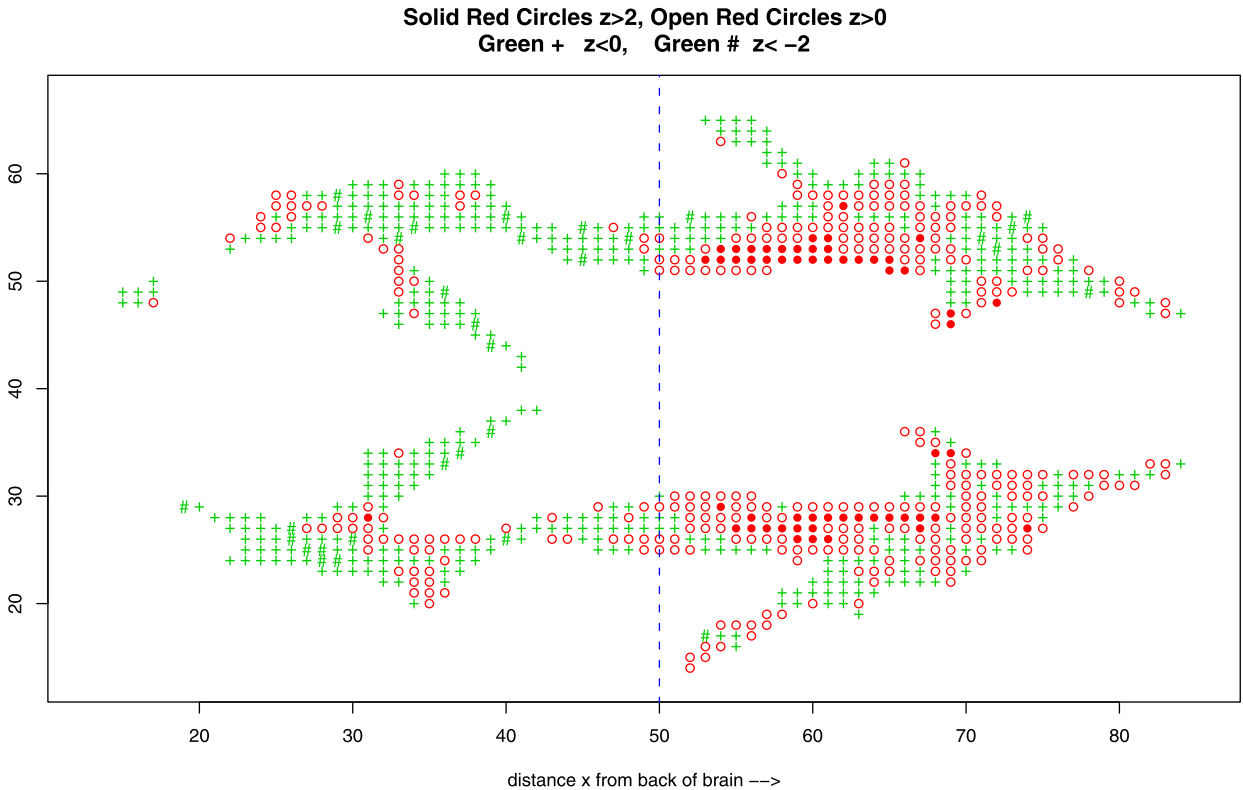}

\caption{DTI study $z$-values comparing 6 dyslexic children with 6
  normal controls, at $N=15\mbox{,}443$ voxels; shown is horizontal section of
  848 voxels; $x$ indicates distance from back of brain (left) to
  front (right). The vertical line at $x=50$ divides the brain into
  back and front halves.}
\label{fig6}
\end{figure*}

At this point one could maintain faith in the theoretical null but at
the expense of concluding that about 2500 (35\%) of the genes are
involved in\break AML/ALL differences. On the other hand, there are plenty
of reasons to doubt the theoretical null. In particular, the leukemia
data comes from an observational study, not a randomized experiment,
so that unobserved covariates (age, sex, health status, race, etc.)
could easily add a component of variance to both the null and non-null
$z$-values.

The crucial question here has\vspace*{1pt} to do with the numerator $E_0(c)$ in
$\hFdr(c)=E_0(c)/N(c)$, the expected number of null cases exceeding
$c$. The theoretical $\n(0,1)$ null predicts many fewer of these than
does the empirical null \eqref{11}. The fact that we might estimate
the appropriate null distribution from evidence at hand---bordering
on heresy from the point of view of classical testing theory---shows
the opportunities inherent in large-scale studies, as well as the
novel inferential questions surrounding the use of indirect evidence.

\section{Relevance}\label{sec8}

Large-scale testing algorithms are usually carried out under the tacit
assumption that all available cases should be analyzed together: for
instance, employing a single false discovery analysis for all the
genes in a given microarray experiment. This can be a dangerous
assumption, as the example illustrated in Figure \ref{fig6} will show.

Twelve children, six dyslexics and six normal controls, received DTI
(diffusion tensor imaging) scans, measuring fluid diffusion at
$N=15\mbox{,}443$ locations (voxels) in the brain. A $z$-value $z_i$ was
computed at each voxel such that the theoretical null hypothesis
$z_i\sim\n(0,1)$ should apply to locations where there is no
dyslexic/normal distributional difference. The goal of course was to
pinpoint areas of genuine difference.

\begin{figure*}

\includegraphics{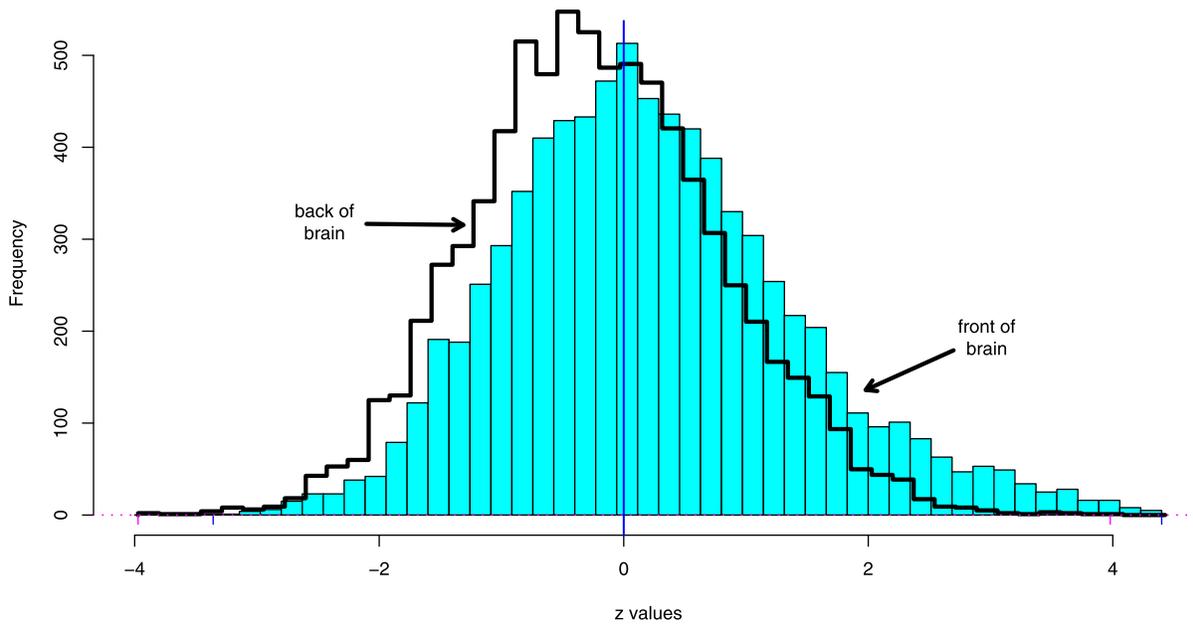}

\caption{Separate histograms for $z_i$'s from the front and back
  halves of the brain, DTI study. The heavy right tail of the
  front-half data yields 281 significant voxels in an Fdr test,
  control level $q=0.10$.}\label{fig7}
\end{figure*}

Figure \ref{fig6} indicates the $z$-values in a horizontal slice of the brain
about half-way from bottom to top. Open circles, colored red, indicate
$z_i\geq0$, solid red circles $z_i\geq2$; green + symbols indicate
$z_i<0$, with green \# for $z_i<-2$. The $x$-axis measures distance from
the back of the brain to the front, left to right.

Spatial correlation among the $z_i$'s is evident: red circles are near
red circles and green +'s near other green +'s. The
Benjamini--Hochberg Fdr control algorithm tends to perform as claimed
as an hypothesis-testing device, even under substantial correlation.
However, there is an empirical Bayes price to pay: correlation makes
$\hFdr(c)$ \eqref{10} less dependable as an estimate of the true Bayes
probability \eqref{9}. Just how much less is a matter of current
study.

There is something else to worry about in Figure \ref{fig6}: the front half
of the brain, $x\geq50$, seems to be redder (i.e., with more positive
$z$-values) than the back half. This is confirmed by the superimposed
histograms for the two halves, about 7700 voxels each, seen in
Figure \ref{fig7}. Separate Fdr tests at control level $q=0.10$ yield 281
``significant'' voxels for the front-half data, all those with
$z_i\geq2.69$, and none at all for the back half. But if we analyze
all 15443 voxels at once, the Fdr test yields only 198 significant
voxels, those having $z_i\geq3.02$. Which analysis is correct?

This is the kind of question my warning about difficult new inference
problems was aimed at. Notice that the two histograms differ near
their centers as well as in the tails. The Fdr analyses employed
thoretical $\n(0,1)$ null distributions. Using empirical nulls as with
the leukemia data gives quite different null distributions, raising
further questions about proper comparisons.

\begin{figure*}

\includegraphics{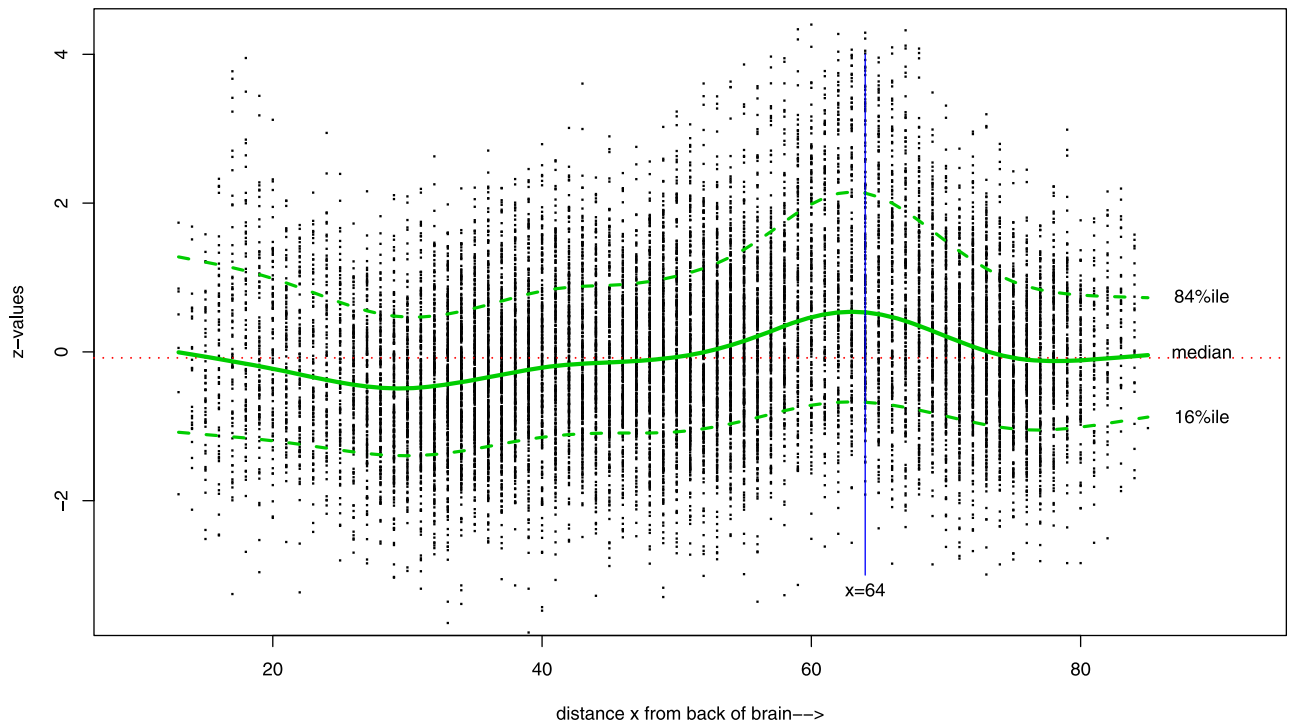}

\caption{$z$-values for the 15,443 voxels plotted versus their
  distance from the back of the brain. A disturbing wave pattern is
  evident, cresting near $x=64$. Most of the 281 significant voxels in
  Figure \protect\ref{fig7} come from this crest.}
\label{fig8}
\end{figure*}

The front/back division of the brain was arbitrary and not founded on
any scientific criteria. Figure \ref{fig8} shows all 15,443 $z_i$'s plotted
against $x_i$, the voxel's distance from the back. We see waves in the
$z$-values, at the lower percentiles as well as at the top, cresting
near $x=64$. Disturbingly, most of the 281 significant voxels for the
front-half analysis came from this crest.

Maybe I should be doing local Fdr tests of some sort, or perhaps making
regression adjustments (e.g., subtracting off the  running median)
before applying an Fdr procedure. We have returned to a version of the
Clemente problem: which are the relevant voxels for deciding whether or
not any given voxel is responding differently in dyslexics and
controls? In other words, where is the relevant indirect information?

\section{The Normal Hierarchical Model}\label{sec9}

My final example of indirect evidence and empirical Bayes inference
concerns the \textit{normal hierarchical model}. This is a simple but
important Bayesian model where $\mu$, a parameter of interest, comes
from some prior density $g(\cdot)$ and we get to observe a normal
variate $z$ centered at $\mu$,
\begin{equation}
\mu\sim g(\cdot)\quad\mbox{and}\quad z|\mu\sim\n(\mu,1).
\label{12}
\end{equation}
Both the James--Stein and Benjamini--Hochberg estimators can be
motivated from \eqref{12},
\begin{eqnarray}\label{13}
\mathrm{JS}\dvtx g&=&\n(M,A)\quad\mbox{and}\nonumber\\[-8pt]\\[-8pt]
\mathrm{BH}\dvtx g&=&p_0\delta_0+p_1g_1.\nonumber
\end{eqnarray}
In the latter, $\delta_0$ is a delta function at 0 while $g_1$ is an
arbitrary density giving $f_1$ in \eqref{7} by convolution,
$f_1=g_1*\varphi$ where $\varphi$ is the standard normal density.

In the BH setting, we might call $\mu_i$ (the value of $\mu$ for the
$i$th case) the \textit{effect size}. For prediction purposes, we want
to identify cases not only with $\mu_i\neq0$ but with large effect
size. A very useful property of the normal hierarchical model
\eqref{12} allows us to calculate the Bayes estimate of effect size
directly from the convolution density $f=g*\varphi$ without having to
calculate $g$,
\begin{equation}
f(z)=\int_{-\infty}^\infty\varphi(z-\mu)g(\mu)\,d\mu.
\label{14}
\end{equation}
\begin{lemma}\label{lem1}
Under the normal hierarchical model \eqref{12},
\begin{equation}
E\{\mu|z\}=z+f'(z)/f(z),
\label{15}
\end{equation}
where $f'(z)=df(z)/dz$.
\end{lemma}

\begin{figure*}

\includegraphics{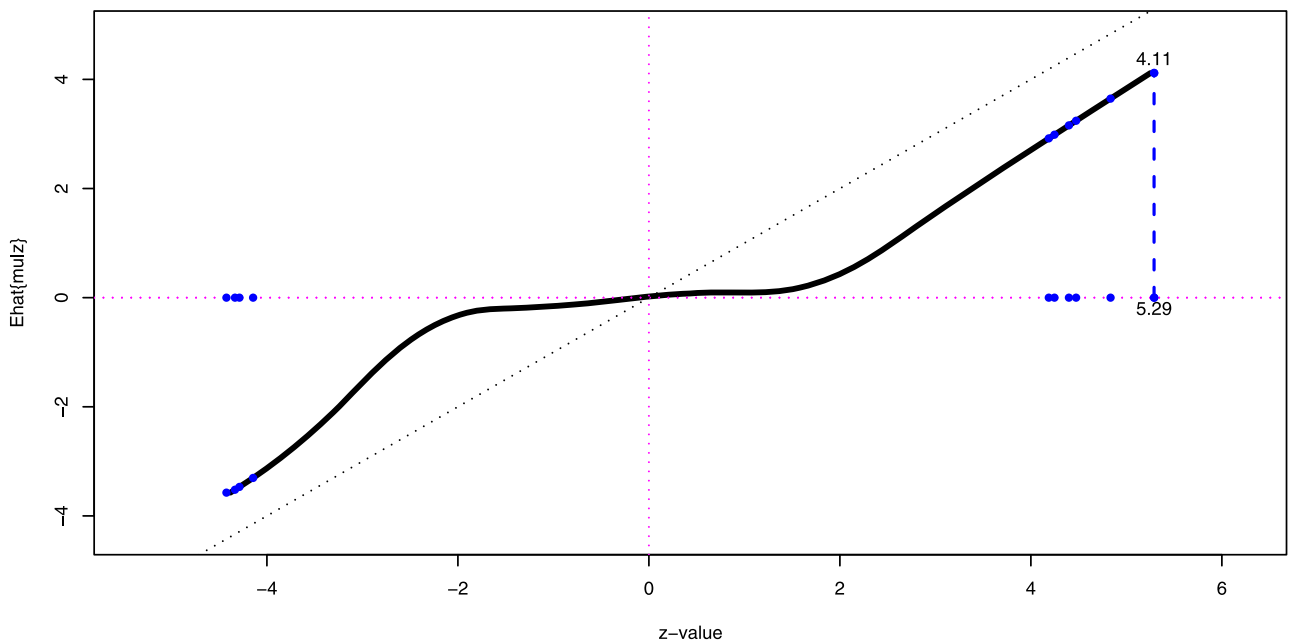}

\caption{Empirical Bayes effect size estimate $\hat{E}\{\mu|z\}$
  (\protect\ref{16}), prostate data of Figure \protect\ref{fig3}. Dots indicate the top 10
  genes, those with the greatest values of $|z_i|$. The top gene,
  $i=610$, has $z_i=5.29$ and estimated effect size $4.11$.}
\label{fig9}
\end{figure*}

The marginal density of $z$ in model \eqref{12} is $f(z)$. So if we
observe $\mathbf{z}=(z_1,z_2,\dots,z_N)$ from repeated realizations of
$(\mu_i,z_i)$, we can fit a smooth density estimate $\hat{f}(z)$ to the
$z_i$'s and use the lemma to approximate $E\{\mu_i|z_i\}$,
\begin{equation}\label{16}
\qquad\mathbf{z}\longrightarrow\hat{f}(z)\longrightarrow\hat{E}\{\mu_i|z_i\}=z_i+\hat{f}'(z_i)\big/f(z_i).
\end{equation}
This has been done in Figure \ref{fig9} for the prostate data of Figure \ref{fig3},
with $\hat{f}(z)$ a natural spline, fit with 7 degrees of freedom to
the heights of Figure \ref{fig3}'s histogram bars (all of them, not just the
central ones we used to estimate empirical nulls).

The effect size estimates $\hat\mu_i=\hat{E}\{\mu_i|z_i\}$ are nearly
zero for $|z_i|$ less than 2 but increase linearly outside of this
interval. Gene 610 has the largest $z$-value, $z_{610}=5.29$, with
estimated effect size $\hat\mu_{610}=4.11$. Table \ref{table2} shows the top
10 genes in order of $|z_i|$, and their corresponding effect sizes
$\hat\mu_i$. The $\hat\mu_i$ values are shrunk toward the origin, but
in a manner appropriate to the BH prior in \eqref{13}, not JS.

The necessity for shrinkage reflects \textit{selection bias}: the top
10 genes were winners in a competition with 6023 others; in addition
to being ``good'' in the sense of having genuinely large effect sizes,
they've probably been ``lucky'' in that their random measurement
errors were directed away from zero. Regression to the mean is another
name for the shrinkage effect.

A wonderful fact is that Bayes estimates are immune to selection bias!
If $\hat\mu_{610}=4.11$ was the actual Bayes estimate
$E\{\mu_{610}|\mathbf{z}\}$ then it would not matter that we became
interested in Gene 610 only after examining all 6033 $z$-values:
$4.11$ would still be our estimate. This may seem surprising, but it
follows immediately from Bayes theorem, a close cousin to results such
as ``Bayes inference in a clinical trial is not affected by
intermediate looks at the data.''

\begin{table}[b]
\caption{Top 10 genes, those with largest values of $|z_i|$, in the
  prostate study and their corresponding effect size estimates $\hat\mu_i$}\label{table2}
\begin{tabular*}{\columnwidth}{@{\extracolsep{\fill}}lcd{2.2}d{2.2}@{}}
\hline
&\textbf{Gene}&\multicolumn{1}{c}{$\bolds{z}$\textbf{-value}}&\multicolumn{1}{c@{}}{$\bolds{\hat\mu_i=\hat{E}\{\mu_i|z_i\}}$}\\
\hline
\phantom{0}1&\phantom{0}610&5.29&4.11\\
\phantom{0}2&1720&4.83&3.65\\
\phantom{0}3&\phantom{0}332&4.47&3.24\\
\phantom{0}4&\phantom{0}364&-4.42&-3.57\\
\phantom{0}5&\phantom{0}914&4.40&3.16\\
\phantom{0}6&3940&-4.33&-3.52\\
\phantom{0}7&4546&-4.29&-3.47\\
\phantom{0}8&1068&4.25&2.99\\
\phantom{0}9&\phantom{0}579&4.19&2.92\\
10&4331&-4.14&-3.30\\
\hline
\end{tabular*}
\end{table}

Any assumption of a Bayes prior is a powerful statement of indirect
evidence. In our example it amounts to saying, ``We have an infinite
number $N$ of relevant prior observations $(\mu,z)$ with $z=5.29$, and
for those the average value of $\mu$ is $4.11$.'' The $N=\infty$ prior
observations outweigh any selection effects in the comparatively puny
current sample, which is another way of stating the wonderful fact.

Of course, we usually don't have an infinite amount of relevant past
experience. Our empirical Bayes estimate $\hat\mu_{610}=4.11$ is based
on just the $N=6033$ observed $z_i$ values. One might ask how immune
are \textit{empirical} Bayes estimates to selection bias? This is the
kind of important indirect-evidence question that I'm hoping
statisticians will soon be able to answer.\looseness=1

\section{Learning From the Experience of Others}\label{sec10}

As I said earlier, current statistical practice is dominated by
frequentist methodology based on direct evidence. I don't believe
this kind of single-problem $N=1$ thinking, even supplemented by
aggressive regression technology, will carry the day in an era of
enormous data sets and large-scale inferences. The proper use of
indirect evidence---learning from the experience of others---is a
pressing challenge for both theoretical and applied statisticians.
Perhaps I should just say that frequentists need to become better
Bayesians.

This doesn't let Bayesians off the hook. A ``theory of everything''
can be a dangerous weapon in the messy world of statistical
applications. The tacit assumption of having $N=\infty$ relevant past
cases available for any observed value of the data can lead to a
certain reckless optimism in one's conclusions. Frequentism is a leaky
philosophy but a good set of work rules. Its fundamentally
conservative attitude encourages a careful examination of what can go
wrong as well as right with statistical procedures and, as I've tried
to say, there's no shortage of wrong steps possible in our new
massive-data environment.

Fisherian procedures, which I haven't talked about here, often
provide a pleasant compromise between Bayesian and frequentist
methodology. Maximum likelihood estimation in particular can be
interpreted from both viewpoints, as a preferred way of combining
evidence from different sources. Fisher's theory was developed in a
small-sample direct-evidence framework, however, and doesn't answer
the questions raised here. Mainly it makes me hope for a new
generation of Fishers, Neymans, Hotellings, etc., to deal with
21st-century problems.

Empirical Bayes methods seem to me to be the most promising candidates
for a combined\break Bayesian/frequentist attack on large-scale data
analysis problems, but they have been ``promising'' for 50-plus years
now, and have yet to form into a coherent theory. Most pressingly,
both frequentists and Bayesians enjoy convincing information theories
saying how well one can do in any given situation, while empirical
Bayesians still operate on an ad hoc basis.

This is an exciting time to be a statistician: we have a new class of
difficult but not impossible problems to wrestle with, which is the
most any intellectual discipline can hope for. The wrestling process
is already well underway, as witnessed in our journals and
conferences. Like most talks that have ``future'' in the title, this
one will probably seem quaint and limited not very long from now, but
perhaps the discussants will have more to say about that.

\section*{Acknowledgments}
Supported in part by NIH Grant 8R01 EB002784 and by NSF Grant DMS-08-04324.

\def\bibname{Relevant References}

\vspace*{-2pt}

\end{document}